\documentclass[letterpaper,english,reprint,aps,prx,showpacs,superscriptaddress,groupedaddress]{revtex4-1}
\usepackage[T1]{fontenc}
\usepackage{CJKutf8}
\usepackage{babel}
\usepackage{verbatim}
\usepackage{amsmath}
\usepackage{amssymb}
\usepackage{graphicx}
\usepackage[unicode=true,bookmarks=false,breaklinks=false,pdfborder={0 0 1},colorlinks=false]{hyperref}
\usepackage{soul}

\makeatletter

\usepackage[usenames,dvipsnames,svgnames,table]{xcolor}
\definecolor{airforceblue}{rgb}{0.36, 0.54, 0.66}

\usepackage{babel}

\hypersetup{
     colorlinks=true,
     linkcolor=blue,
     filecolor=blue,
     citecolor = blue,      
     urlcolor=blue,
     }
     
\makeatother

\begin{document}
\title{Ballistic-to-diffusive transition in spin chains with broken integrability}
\author{João S. Ferreira$^{1}$}
\author{Michele Filippone$^{1}$}
\affiliation{$^{1}$Department of Quantum Matter Physics, University of Geneva, 24 Quai Ernest-Ansermet,
CH-1211 Geneva, Switzerland}
\date{\today}
\begin{abstract}
\noindent %
\noindent\begin{minipage}[t]{1\columnwidth}%
\global\long\def\ket#1{\left| #1\right\rangle }%
\global\long\def\bra#1{\left\langle #1 \right|}%
\global\long\def\kket#1{\left\Vert #1\right\rangle }%
\global\long\def\bbra#1{\left\langle #1\right\Vert }%
\global\long\def\braket#1#2{\left\langle #1\right.\left|#2\right\rangle }%
\global\long\def\bbrakket#1#2{\left\langle #1\right.\left\Vert #2\right\rangle }%
\global\long\def\av#1{\left\langle #1\right\rangle }%
\global\long\def\Tr{\text{Tr}}%
\global\long\def\id{\mathbb{I}}%
\global\long\def\inf{\infty}%
\global\long\def\a{\alpha}%
\global\long\def\b{\beta}%
\global\long\def\c{\hat{c}}%
\global\long\def\g{\gamma}%
\global\long\def\G{\Gamma}%
\global\long\def\w{\omega}%
\global\long\def\O{\Omega}%
\global\long\def\d{\delta}%
\global\long\def\D{\Delta}%
\global\long\def\m{\mu}%
\global\long\def\LL{\hat{\mathcal{L}}}%
\global\long\def\H{\mathcal{H}}%
\global\long\def\s{\sigma}%
\global\long\def\r{\rho}%
\global\long\def\l{\lambda}%
\global\long\def\T{\Theta}%
\global\long\def\dag{\dagger}%
\global\long\def\L{\Lambda}%
\global\long\def\e{\epsilon}%
\global\long\def\x{\chi}%
\global\long\def\fl{V^{2}f_{V^{2}}(L)}%
\global\long\def\gl{\D^{2}f_{\D^{2}}(L)}%
\end{minipage}
We study the ballistic-to-diffusive transition induced by the weak breaking of integrability in a boundary-driven XXZ spin-chain. 
Studying the evolution of the spin current density $\mathcal J^s$ as a function of the system size $L$, we show that, accounting for boundary effects, the transition has a non-trivial universal behavior close to the XX  limit. It is controlled by the scattering length $L^*\propto V^{-2}$, where $V$ is the strength of the integrability breaking term. 
In the XXZ model, the interplay of interactions controls the emergence of a transient ``quasi-ballistic'' regime at length scales much shorter than $L^*$.
This parametrically large regime is characterized by a strong renormalization of the current which forbids a universal scaling, unlike the XX model.
Our results are based on Matrix Product Operator numerical simulations and agree with perturbative analytical calculations.


\end{abstract}
\maketitle

\section{Introduction\label{sec:Introduction}}

A central assumption of statistical mechanics is
that many-body interactions bring isolated out-of-equilibrium systems towards thermal equilibrium 
\cite{Deutsch1991,Srednicki1994,Rigol2008,DAlessio2016}. The phenomenon of thermalization in normal - metallic - conductors is generally associated with diffusion. Globally conserved quantities such as energy, charge, spin  or mass spread uniformly
all over the system according to Fick's law
\begin{align}\label{eq:fick}
\mathcal{J} & =-D \nabla n\,,
\end{align}
in which the diffusion
constant $D$ relates the current density $\mathcal{J}$ to  the application of a density gradient $\nabla n$.
Recently, it has been  observed that in one-dimension, quantum
integrable systems  defy thermalization~\cite{Kinoshita2006}. This discovery has triggered an intense effort to understand the non-trivial dynamics
of such systems under the recently developed framework of generalized
hydrodynamics~\cite{CastroAlvaredo2016,Bertini2016}.
In particular, the presence of an extensive amount of conservation laws in integrable systems~\cite{Rigol2007} generically leads to ballistic transport of conserved quantities~\cite{Ilievski2017}. An important case is spin transport in the XXZ model, which can  show, for some choice of the model parameters,  (super)diffusive behavior~\cite{ljubotina2019kardar,Gopalakrishnan2019,Prosen2011,Ljubotina2017,Bertini2020,Ilievski2018,DeNardis2020,DeNardis2018}. 

Unavoidable deviations from the realization of a perfect, fine-tuned integrable system lead to integrability breaking (IB). In that case, one typically expects the slow establishment of a chaotic-diffusive regime on time scales given by Fermi's golden rule (FGR)
\cite{Tang2018,Mallayya2019}.
Nevertheless, the investigation on how IB triggers proper diffusive regimes for transport remains at a preliminary stage.  Even though recent works~\cite{Friedman2020,Durnin2020,Moller2020}
derived a generalized expression of FGR to
describe diffusive hydrodynamics caused by integrability breaking, the
onset of diffusion may still unveil highly non-trivial behavior~\cite{Jung2006,Bulchandani2020}.
Additionally, the onset of chaotic/diffusive behavior, for fixed 
weak interactions is not controlled by Fermi's golden rule at small system sizes~\cite{Silvestrov1998,Neuenhahn2012,Pandey2020}.
Recent works have also pointed out that the emergence of chaotic/diffusive behavior may not be fully related to  the usual measurements
of quantum chaos, such as level repulsion~\cite{Brenes2018,Brenes2020,Brenes2020b} or the eigenstate thermalization hypothesis~\cite{Luitz2016}. 

Quantum quenches are a very efficient and widespread protocol used to study the relaxation dynamics of such many-body systems~\cite{Essler2016,Calabrese2016,Bernard2016,DeNardis2018,Biella2019,VonKeyserlingk2018,Alba2019,Jesenko2011,Ilievski2015}. They are regularly performed in state-of-the-art cold-atom experiments~\cite{Trotzky2012,Gring2012,Hofferberth2007,Jepsen2020} and can be  efficiently simulated by  numerical approaches~\cite{Schollwoeck2005,White2009,White1992,PerezGarcia2007,Kennes2016}. 
Nevertheless,  the description of the long-time dynamics driven by weak IB remains  challenging for the available analytical and numerical studies.

\begin{figure}[b]
\includegraphics[width=1\columnwidth]{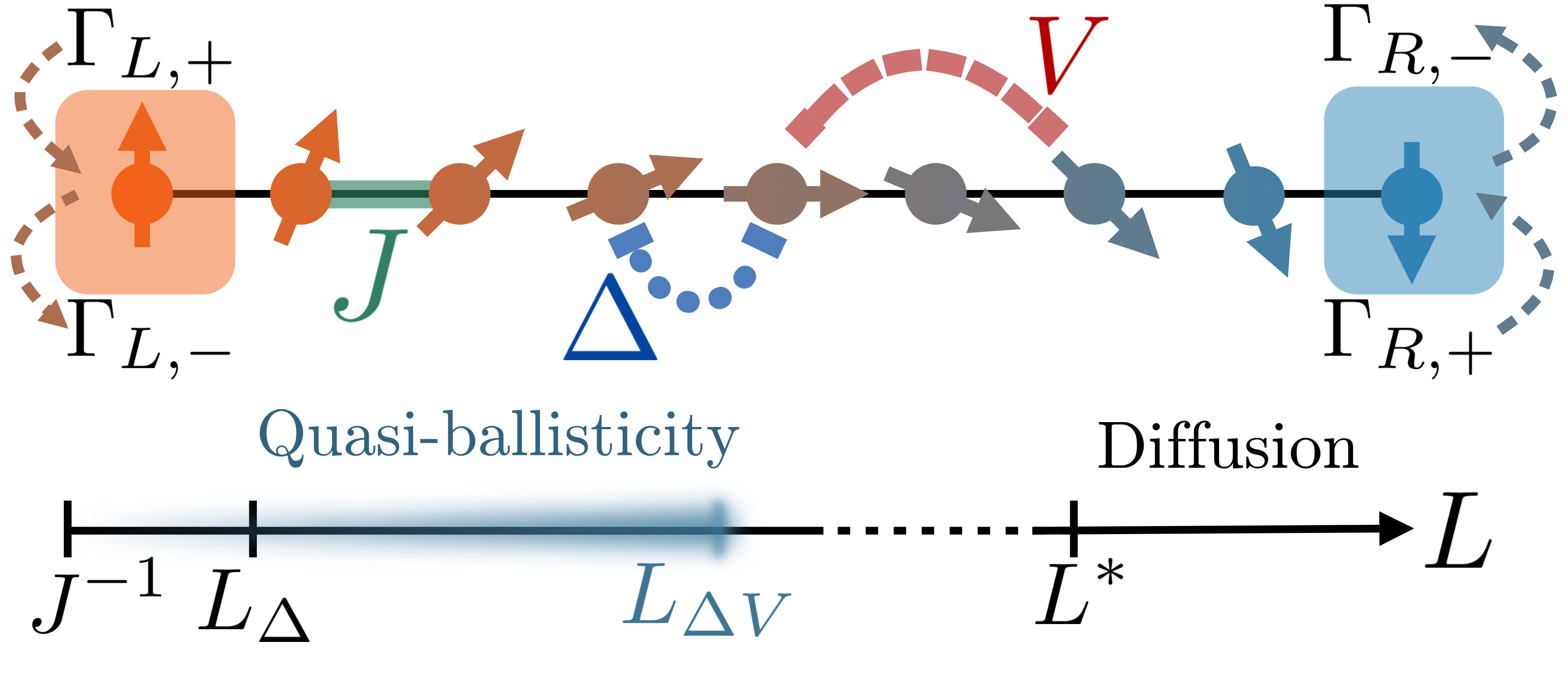} \caption{ Top) System under study:  a spin current is induced  via  biased  Lindblad jump operators at the edge of a XXZ spin chain described by Eq.~\eqref{eq:xxz} and~\eqref{eq:nn}. Bottom) Schematic behavior of the steady-state current as a function of the system size $L$. The ballistic (size-independent) regime in the XXZ model sets in after the length scale $L_\D$. Breaking integrability triggers diffusion beyond the scattering length $L^*$, given by Fermi's golden rule.  The ballistic-to-diffusive crossover regime is controlled by the emerging length scale $L_{\D V}\ll L^*$, which defines a parametrically large ``quasi-ballistic'' regime.   }
\label{fig:setup} 
\end{figure}

We chose to address this issue from a different but complementary point of view. We investigate at which system {\it sizes} do weak IB interactions of strength $V$ lead to a ballistic-to-diffusive transition in one-dimensional spin-chains. $V$ is compared to the spin-exchange strength $J$. We  study the effect of next-to-nearest neighbor interactions on the stationary current carried by a ballistic XXZ spin-chain driven at its boundaries, see Fig.~\ref{fig:setup}. The boundary terms induce a bias in the magnetization that in turn generates a spin current density, $\mathcal{J}^s$. This approach has the advantage to directly probe the stationary properties of highly excited many-body systems with large system sizes~\cite{Ljubotina2017,Znidaric2019} and it has been recently formulated in terms of the local properties of the interacting region~\cite{jin_generic_2020}. 
Different transport regimes are characterized by a unique scaling of current density with the system size $L$, $\mathcal{J}^s \propto L^{-\a}$. Here, we focus on the crossover from ballistic ($\a=0$) to diffusive ($\a=1$) as we approach the thermodynamic limit.

The presence of non-integrable interactions in the Hamiltonian introduces a natural length scale to the problem: the scattering length $L^*\propto V^{-2}$, as suggested from perturbation theory and FGR. One could thus expect that the current  scales as a universal function of $L/L^*$. 

In this work, we show that the observation of such scaling is not trivial in the case of the XX chain in the presence of  IB perturbations. Our numerical calculations show that boundary effects strongly affect the ballistic-to-diffusive transition at short length scales, which are not controlled by the scattering length $L^*$. Nevertheless, relying on perturbation theory, we derive a non-trivial universal scaling of the current on the system size $L$. This scaling accounts for boundary corrections and it allows an accurate extrapolation of the results to the thermodynamic  and $V\rightarrow 0$ limit. Our results are consistent with the establishment of a universal scaling as a function of $L/L^*$ in this limit and also allow a good estimate of the diffusion constant $D$.  


We then extend to the study of the integrable ballistic case in the presence of interactions ($|\Delta|<1$).  In this case, we show  the emergence of \textit{linear} corrections to the current in the IB strength $V$. These corrections control the  emergence of a mesoscopic ``quasi-ballistic'' regime, in which the ballistic current is strongly renormalized before the onset of diffusion, see Fig.~\ref{fig:setup}. More specifically, we show that, up to parametrically large systems $L< L_{\D V}\propto  V^{-1}<L^*$, IB does not lead to current suppression. Instead the ballistic current is strongly renormalized and, for repulsive interactions $(V>0)$, it may even increase with respect to the integrable case.

The results in this paper are expected to manifest themselves in real experiments probing the transport~\cite{Krinner2015,Brantut2013,Lebrat2018} and relaxation properties of isolated interacting systems, close to integrable points.


Our paper is structured as follows. In Section~\ref{sec:modelmethods}, we present the system, the Lindblad formalism, which allows to describe a stationary state carrying a current and the numerical approach based on tDMRG. In Section~\ref{subsec:XX-Limit}, we  discuss the universal scaling induced by IB  when perturbing the XX limit. Section~\ref{subsec:XXZ--Ballistic} discusses the effect of  IB in the XXZ model.
Section~\ref{sec:conclusions} is devoted to the discussions of our results and conclusions. The appendices incorporate  details about the tDMRG implementation, perturbation theory and complementary plots to our numerical analysis.


\section{Model and methods}\label{sec:modelmethods} 

We consider the anisotropic Heisenberg (XXZ) chain in one dimension~\cite{Bethe1931}
\begin{equation}\label{eq:xxz}
H_{\text{XXZ}}=J\sum_{i=1}^{L-1}(\s_{i}^{x}\s_{i+1}^{x}+\s_{i}^{y}\s_{i+1}^{y})+\Delta\sum_{i=1}^{L-1}\s_{i}^{z}\s_{i+1}^{z}\,,
\end{equation}
in which $\sigma^{x,y,z}$ are the standard Pauli matrices and $L$
the number of spins in the system. We set $J=1$. The model~\eqref{eq:xxz} is integrable and
its ground state is gapless for $\Delta\in[-1,1]$ and gapped otherwise.
Remarkably, it supports ballistic spin transport at finite energy
density in the gapless phase~\cite{Prosen2011}, superdiffusion
at the isotropic point $|\Delta|=1$
and normal diffusion otherwise~\cite{Ljubotina2017,Gopalakrishnan2019}.

Transition to a diffusive regime is expected when breaking integrability. For the remainder of the paper, we explicitly break integrability by adding global next-to-nearest neighbor (NNN) interactions
\begin{equation}\label{eq:nn}
H_{\rm NNN}=V\sum_{i=1}^{L-2}\s_{i}^{z}\s_{i+2}^{z}\,.
\end{equation}
We recall that $V$ has units of the $J$ coupling and, for the remainder of the paper, we consider weak ($V\ll 1$) and moderate ($V\approx 0.5$) interactions.

To study transport, we numerically mimic the experimentally relevant
situation~\cite{Krinner2015,Brantut2013,Lebrat2018} in which the system is coupled at its two ends to a left
(L) and a right (R) magnetization reservoir. If there is a small magnetization bias, it induces a non-equilibrium steady state (NESS) carrying spin current, see Fig. \ref{fig:setup}. Coupling to external (Markovian) reservoirs results in a non-unitary evolution of the system's density matrix $\rho$. We simulate this evolution with the Lindblad master
equation $d\rho/dt=\LL(\rho)$~\cite{Gorini1976,Lindblad1976}, where $\LL$ is the Liouvillian superoperator which describes the non-unitary dynamics of the system
\begin{align}
\LL(\rho)=-i[H,\rho]+\sum_{\substack{\alpha=L,R\\
\tau=\pm}
}2\Gamma_{\alpha\tau}\rho\Gamma_{\a\tau}^{\dagger}-\left\{ \rho,\Gamma_{\a\tau}^{\dagger}\Gamma_{\a\tau}\right\} \,. \label{eq:master-eq}
\end{align}
The dissipative dynamics induced by the reservoirs is
expressed in terms of the jump operators $\Gamma_{\a\tau}=\sqrt{\gamma(1+\tau\mu_{\alpha})}\sigma_{\a}^{\tau}$,
where $\gamma$ is the injection/loss rate and $\mu_{L}=-\mu_{R}=\frac{\d\mu}{2}$,
with $\delta\mu$ being the bias in magnetization, see also Fig.~\ref{fig:setup}. To simplify the expressions, we fix $\gamma=1$. Equation~\eqref{eq:master-eq} effectively describes a system attached to weakly magnetized reservoirs which have a temperature much larger than the energy spectrum of the system~\cite{BreuerPetruccione_book,GardinerZoller_quantumnoise}. This notion has  been recently put on solid grounds in Ref.~\cite{jin_generic_2020}.

For small magnetization bias, $|\d\mu|\ll1$, the
NESS induced by Eq.~\eqref{eq:master-eq}
is close to  $\rho_\infty=\id^{\otimes L}/2^L$ \cite{Prosen2009}. It  describes the infinite temperature situation in which, irrespective of the system Hamiltonian, each spin is in a classical mixed state with the same 1/2 probability to be up or down.
The stationary state carries a non-zero average spin-current density $\mathcal{J}^{s}=2\sum_{i}^{L} \left< \sigma_{i}^{+}\sigma_{i+1}^{-}-\sigma_{i}^{-}\sigma_{i+1}^{+} \right>/L$. 

The biased jump operators $\Gamma_{\alpha\tau}$ enforce different spin densities at the two ends of the chain~\footnote{If the boundary spins that are connected to the jump operators were to be isolated from the chain's bulk, their occupation would be $\mu_\alpha\leq1$.} and allow for a direct investigation of the spin-current. In particular, the dependence of the spin current $\mathcal{J}^{s}$ as a function of system size $L$ allows us to distinguish between ballistic and diffusive transport regimes. Ballistic regimes are not described by Fick's law~\eqref{eq:fick} and they are characterized by a steady-state current that does not decay with system size $L$, whereas diffusive regimes are signaled by a current which decays inversely with $L$.

\subsection{Numerical methods}\label{sec:numeric} 

To find the steady-state of the master equation~\eqref{eq:master-eq}, we employ a time-dependent density matrix renormalization-group
(tDMRG) method~\cite{Schollwoeck2005}, implemented with the ITensor library~\cite{itensor}. For $\delta\mu=0$, the steady-state of Eq.~\eqref{eq:master-eq} is the infinite temperature state. We thus perform a real-time evolution of an initial density matrix $\rho(t=0)=\rho_\infty=\mathbb{I}^{\otimes L}/2^L$, which is written in a matrix product operator (MPO) form.
Since the non-equilibrium steady state, $\rho_{\rm NESS}=\lim_{t\rightarrow\inf}\r(t)$, is unique, it is well approximated by $\rho(t)$ for very large times and increasing bond dimensions. By numerically verifying convergence both in time and bond dimension, we are able to compute the NESS for system sizes up to one-hundred sites ($L=100$). Our numerical simulations were carried out for a magnetization bias $\d\mu=0.1$, for which we verified that  the current's response is linear in $\d\mu$. The bond dimension is limited to $\chi=160$ and the time step of the Trotter decomposition ranges from $dt=0.05$ to $dt=0.2$. The interested reader is redirected to App.  \ref{sec:Numerical-details}, where we provide all the necessary details concerning our numerical simulations.

\subsection{Analytical methods}\label{sec:analytic} 
To gain insight in the numerical results, we also rely on perturbation theory to compute the corrections to the spin current caused by weak interactions $V,\Delta\ll J$ in finite-sized systems. Similarly to conventional perturbation theory in the Hamiltonian language, the starting point is a fully diagonalized model. In our case, the reference model is the XX chain with boundary-driving, which is a quadratic model and has been analytically solved relying on the third-quantization formalism~\cite{Prosen2008,Guo2017}. The description of this formalism is rather technical and does not provide special physical insight. We give  thus in Appendix~\ref{app:3quant} all the necessary details and describe here only the main steps. The procedure requires first to map the XX chain onto non-interacting fermions  via the Jordan-Wigner transformation.  In the absence of interactions ($\D=V=0$),  the generator of the dynamics, $\LL_{\rm XX}$, is quadratic in terms of $L$ fermionic annihilation and creation operators. Nevertheless, the fact that the Liouvillian is a non-Hermitian superoperator acting on the density matrix, requires to work on an extended ``third-quantization'' basis  of $2L\times 2L$ super-operators $\{\c_i,\c'_i\}$, which allows to cast the Liouvillian in the diagonal form 
\begin{equation}
\LL_{\text{XX}}(\circ)=\sum_{i}^{2L}\alpha_i\hat{c}'_i\hat{c}_i(\circ)\,,
\end{equation}
in which the spectrum $\{\alpha_i\}$ can be calculated numerically, as detailed in Appendix~\ref{app:3quant}.   In  such basis, the NESS is expressed as a ``vacuum'' state $\r_0$, for which $\c_i(\r_0)=0$,  and the eigenstates of $\LL_{\text{XX}}$ can be constructed from excitations on the vacuum state, $\rho_{\mu}=\sum_{\left\{ \mu_{i}\right\} }\c'_{1}{}^{\mu_{1}}...\c'_{2L}{}^{\mu_{2L}}(\r_0)$.

The goal is to find a perturbative solution to the NESS of Eq.~\eqref{eq:master-eq}, in the form $\rho_{ss}=\sum_{m,n=0}^{\infty}V^{m}\D^{n}\r_{m,n}$. We plug the pertubative ansatz in the  steady-state condition, $\LL(\r_{ss})=0$, and solve it order by order to find~\cite{Li2014} 
\begin{align}
\begin{split} \label{order-exp}
V^{m}\D^{n}\rho_{m,n}=i\LL_{\text{\rm XX}}^{+}\Big(&\big[H_{\text{NNN}},\r^{(m-1,n)}\big] \\
&+\big[H^{J=0}_{\text{XXZ}},\r^{(m,n-1)}\big]\Big)\,,
\end{split}
\end{align}
where we introduced  the Moore-Penrose pseudoinverse of the Liouvillian of the boundary-driven XX chain, $\LL_{\rm XX }(\circ)=\sum_{\alpha_i\neq0}\alpha_i^{-1}\hat{c}'_i\hat{c}_i(\circ)$. Using the third-quantization formalism we thus  find semi-analytic expressions of $\rho_{ss}$ and $\mathcal{J}^s $ up to second order in the interactions $V$ and $\Delta$, which are given in Appendix~\ref{app:3quant}.

\section{IB and  XX Model ($\protect\D=0$)\\The universal crossover to diffusion}\label{subsec:XX-Limit} 

For the XX chain ($\D=V=0$), the MPO expression of the steady-state of Eq.~\eqref{eq:master-eq} has been derived in Refs.~\cite{Znidaric2010,Znidaric2010b}, and found to carry a ballistic spin current $\mathcal J^s=\delta\mu$~\footnote{The general expression for different injections  rates $\gamma_{L/R}\neq1$ on the left and on the right of the chain reads $\mathcal J^s=\delta\mu\cdot  4\gamma_L\gamma_R/[(1+\gamma_L\gamma_R)(\gamma_L+\gamma_R)]$~\cite{Znidaric2010b}.}. Interactions such as Eq.~\eqref{eq:nn} induce inelastic scattering
among free-particles, which leads to a decay of the spin current and
the onset of diffusion in the thermodynamic limit.

\begin{figure}[t]
\begin{centering}
\includegraphics[width=1\columnwidth]{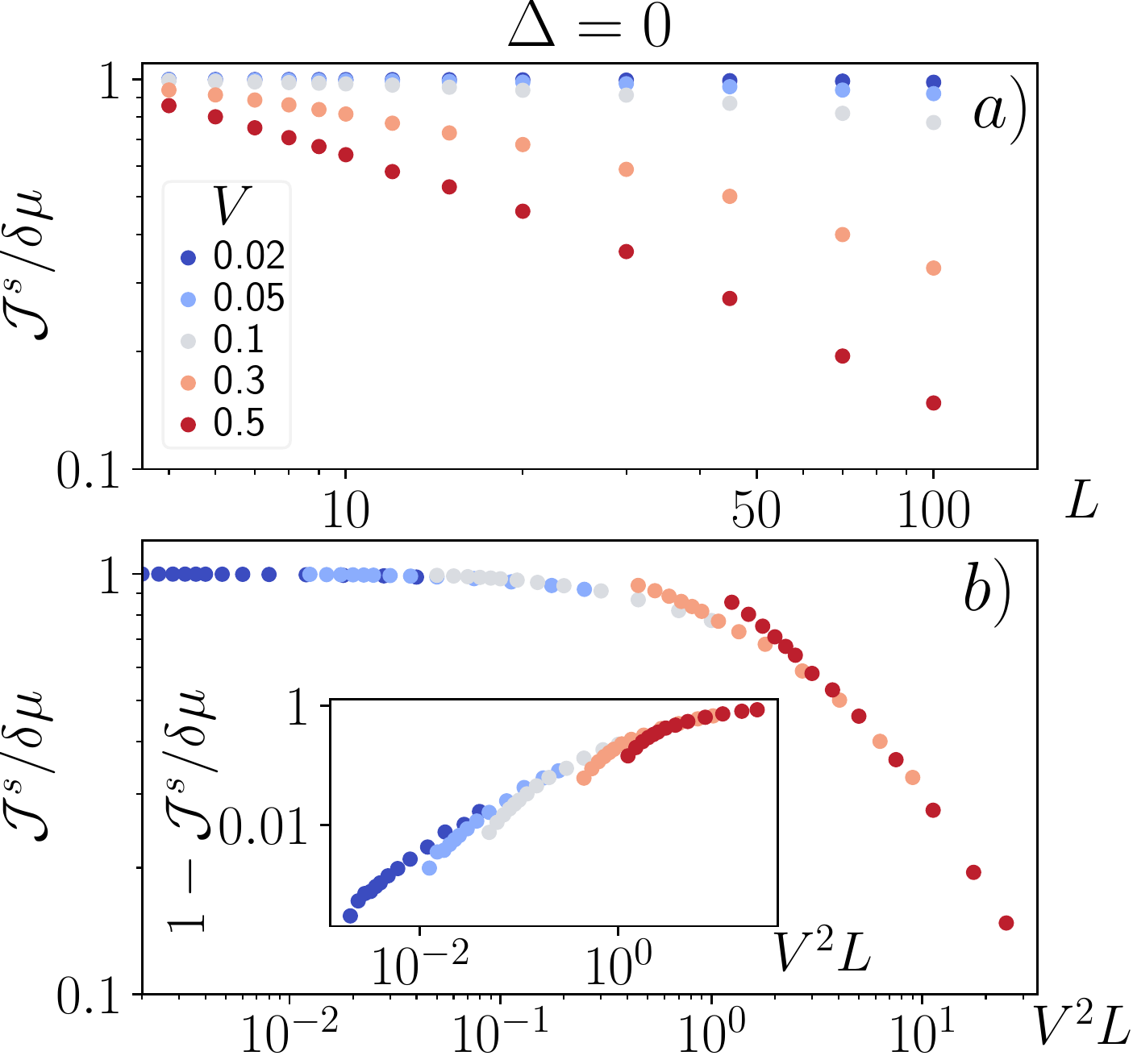} 
\par\end{centering}
\caption{\label{fig:first}Dependence of $\mathcal{J}^{s}$ in the
XX model for different strengths $V$ of the next-nearest neighbors
interaction~\eqref{eq:nn} as a function of: a) system size $L$ and b) the scaling variable $L/L^*$. In the inset, the deviations from the value of the current in the ballistic limit are shown. 
Deviations from an universal scaling are observed in b) for $L\lesssim 10$ at all strengths $V$.}
\end{figure}

For finite but large systems, the ballistic-to-diffusive transition is marked by a sizable deviation from $\mathcal J^s=\delta\mu$ at a crossover length scale $L^*$. According to FGR, this scattering length is expected to  scale as $L^*\sim V^{-2}$ in the $V\rightarrow 0$ limit. 

We numerically compute the current, as a function of the system size $L$, for different strengths $V<J$ of the NNN interaction, Eq.~\eqref{eq:nn}, see Fig.~\ref{fig:first}a.
As expected, with increasing strength of the IB parameter $V$,
the current decreases monotonically for a fixed length $L$
and diffusion sets in at smaller $L$.

Assuming that the only characteristic length (beyond lattice spacing) is the scattering length, it is natural to expect a scaling hypothesis controlled by $L^*$, i.e. $\mathcal J^s/\delta\mu=\mathcal F (L/L^*)=\mathcal F (LV^2)$. However, such scaling ansatz does not allow a perfect collapse of all the numerical curves onto a unique function, see Fig.~\ref{fig:first}b and inset. 
We observe that for small systems, typically up to $L\sim 10$ sites, the current deviates significantly from the scaling for any value of $V$.
The absence of a universal scaling is intriguing and also hinders the possibility to  extrapolate numerical data to arbitrarily large system sizes. It is thus important to understand the deviations and possibly correct them.
\begin{figure}[t]
\begin{centering}
\includegraphics[width=1\columnwidth]{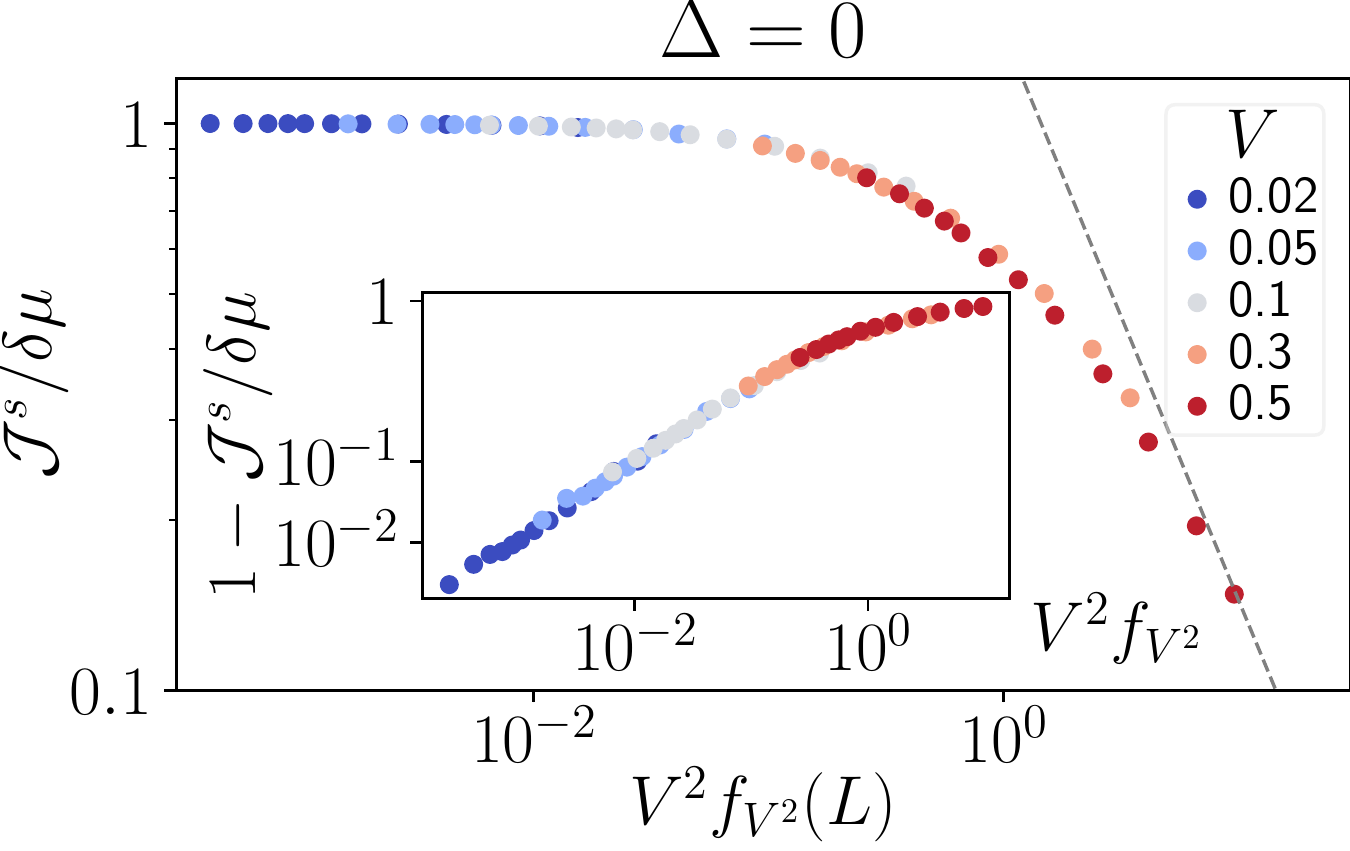} 
\par\end{centering}
\caption{\label{fig:new}Universal scaling dependence of $\mathcal{J}^{s}$ in the XX model for different strengths $V$ of the next-nearest neighbors
interaction~\eqref{eq:nn} as a function $\fl$. The dashed gray line correspond to  the ansatz~\eqref{eq:scalingxx}, valid  for the diffusive regime. In the inset, the deviations from the value of the  current in the ballistic limit are shown. }
\end{figure}

Thus, we rely on perturbation theory, described in Section~\ref{sec:analytic}, to derive the correction to the current at finite sizes $L$. We find that the leading corrections to the current read
\begin{equation}\label{eq:scaling_xx_small}
\mathcal{J}^{s}=\left[1-V^{2}f_{V^{2}}(L)+\mathcal O(V^4)\right]\delta\mu\,.
\end{equation}
Remarkably, the leading corrections to the current have a non-trivial dependence on the system size $L$, through the function $f_{V^2}(L)$, which is plotted in Fig.~\ref{fig:fi} in App.~\ref{sec:perturbation}. Apart from irrelevant corrections for short system sizes,  $f_{V^2}$ is as a linear function with a non-zero offset with respect to the origin, $f_{V^{2}}(L)\approx0.40(L-3.38)$. The divergence of the correction of order $V^2$ in Eq.~\eqref{eq:scaling_xx_small} signals the transition to the diffusive regime in which the current is expected to scale as $L^{-1}$.

Surprisingly, a universal scaling of the numerical data is obtained  when plotting the current as a function of  the non-trivial parameter $\fl$, see Fig.~\ref{fig:new} and inset. The collapse of the curves is excellent up to large system sizes and up to moderate IB strengths, $V=0.5$. This shows the importance of boundary effects in the ballistic-to-diffusive transition triggered by generic interaction on the XX model. The boundary corrections are encoded in the fact that the function $f_{V^2}(L)$ has an offset with respect to a straight line crossing the origin. Such offset becomes negligible for systems sizes $L\geq10$.

The expression~\eqref{eq:scaling_xx_small} is only valid as long as $V^2f_{V^2}(L)\ll1$ and, unluckily, we could not find a good expression fitting the whole curve in Fig.~\ref{fig:new}. Nevertheless, when approaching the diffusive regime, the numerics are consistent with the expression
\begin{equation}\label{eq:scalingxx}
\mathcal{J}^{s}=\frac{1.45}{V^{2}f_{V^{2}}(L)}\d\mu\,,
\end{equation}
corresponding to the gray dashed lines in Fig.~\ref{fig:new}. For asymptotically  large $L$, Eq.~\eqref{eq:scalingxx} acquires the form  $\mathcal{J}^{s}=D^s_{\D=0}\d\mu/2L$, in which $D^s_{\D=0}\approx7.3/V^{2}$ is the spin diffusion constant. This value of the diffusion constant is derived by considering the equivalent of Fick's  law~\eqref{eq:fick} in the spin formulation of the problem, namely $\mathcal{J}^s=-D^s \nabla s^z=D^s\,\delta \mu/2L$, in which $s^{z}=\langle\sigma^{z}\rangle/2$ is the spin expectation value. We have verified numerically,  that $\nabla s^z=-\delta\mu/2L$ gives the correct estimate of the spin-density gradient in diffusive regimes, see  App.~\ref{app:scaling}. One should notice that a precise evaluation of $D^s_{\D=0}$ for weak $V$ would hardly be possible without considering the correct scaling parameter, as it is shown in Fig.~\ref{fig:first}.

This discussion concludes our analysis of the ballistic-to-diffusive transition induced by IB on the XX model.  We showed that   corrections caused by boundary effects affect the scaling of the current for short system sizes. Nevertheless, perturbation theory allows to account for such finite-size corrections and derive a universal ballistic-to-diffusive crossover induced by IB on ballistic, non-interacting regimes.  Our analytical calculation  shows that boundary effects become negligible beyond systems of $L\sim 10$ sites, for which the ballistic-to-diffusive crossover is indeed nicely described  by a universal function of $L/L^*$. It is important to stress that our  analytical calculations are crucial to account for boundary corrections and thus allow an accurate extrapolation of the numerics to the thermodynamic and $V \rightarrow 0$ limit. Without perturbation theory, the universal nature of the scaling would have been difficult to establish based exclusively on numerical data.

We now extend to the interacting and integrable case, showing how  nearest-neighbor interactions, of strength $0<\D<1$, strongly modify the effects of IB on the ballistic regime. 


\section{IB and XXZ model with $|\protect\D|<1$\\}\label{subsec:XXZ--Ballistic}

\subsection{The ballistic, integrable regime}
The sole presence of nearest-neighbor interactions does not hinder
ballistic transport in the thermodynamic limit for $|\Delta|<1$~\cite{Zotos1997,Zotos1999,Prosen2011}. For finite systems, the current depends non-trivially on the system size. For increasing $L$, the current decreases monotonically until it saturates to its ballistic (thermodynamic) value. This saturation occurs beyond a typical length scale $L_{\D}$ which depends on the strength $\D$ of the integrable nearest-neighbor interaction.

\begin{figure}
\begin{centering}
\includegraphics[width=1\columnwidth]{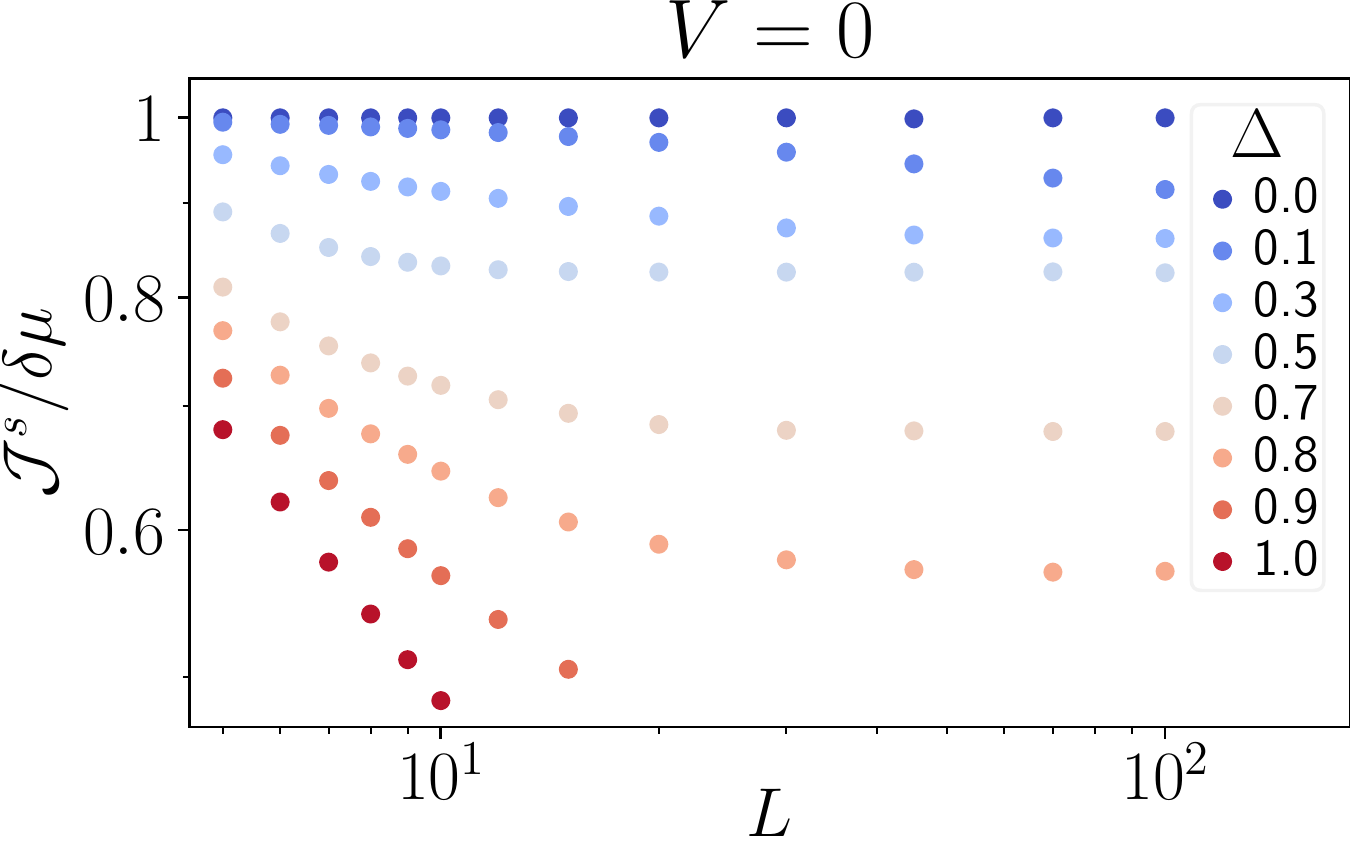} 
\par\end{centering}
\caption{\label{fig:first-2}  Finite size scaling of $\mathcal{J}^{s}$ in
the ballistic regime of the XXZ model~\eqref{eq:xxz}, for different
$0<\protect\D<1$. For $\Delta<1$, the current always saturates to a constant value signaling the ballistic regime.}
\end{figure}

The behavior of $\mathcal J^s$  as function of $L$ is shown in Fig.~\ref{fig:first-2}, which reproduces
the findings of Ref.~\cite{Znidaric2011} and that we display here for clarity. 
To our knowledge, the exact size dependence of the current is unknown.  It is possible to obtain perturbatively the finite-size behavior of the current for $\D\rightarrow 0$
\begin{equation}\label{eq:JsXXZ-PT}
\mathcal J^s=\Big[1-\gl +\mathcal{O}(\D^3) \Big]\delta\mu\,,
\end{equation} 
where $f_{\Delta^2}$ is  a linear function similar to $f_{V^2}$, see App. \ref{sec:perturbation}. In analogy to the discussion of the previous section, Eq.~\eqref{eq:JsXXZ-PT} is only valid for system sizes $L<L_{\D}$, in which, for $\D\ll1$, $L_\D \propto1/\D^2$.  Beyond $L_{\D}$, the perturbative corrections diverge linearly in $L$ and miss the saturation of the current which, to be derived, would require the re-summation of the  perturbation theory in $\D$ to all orders. It should be stressed that, even though the expansions~\eqref{eq:scaling_xx_small} and~\eqref{eq:JsXXZ-PT} look almost identical, their linear divergences do not signal analogous behaviors in the thermodynamic limit. In particular, in the non-integrable case, one would find the  diffusive current suppression described by Eq.~\eqref{eq:scalingxx}.

\subsection{Strong linear effects induced by IB}
\begin{figure}
\begin{centering}
\includegraphics[width=1\columnwidth]{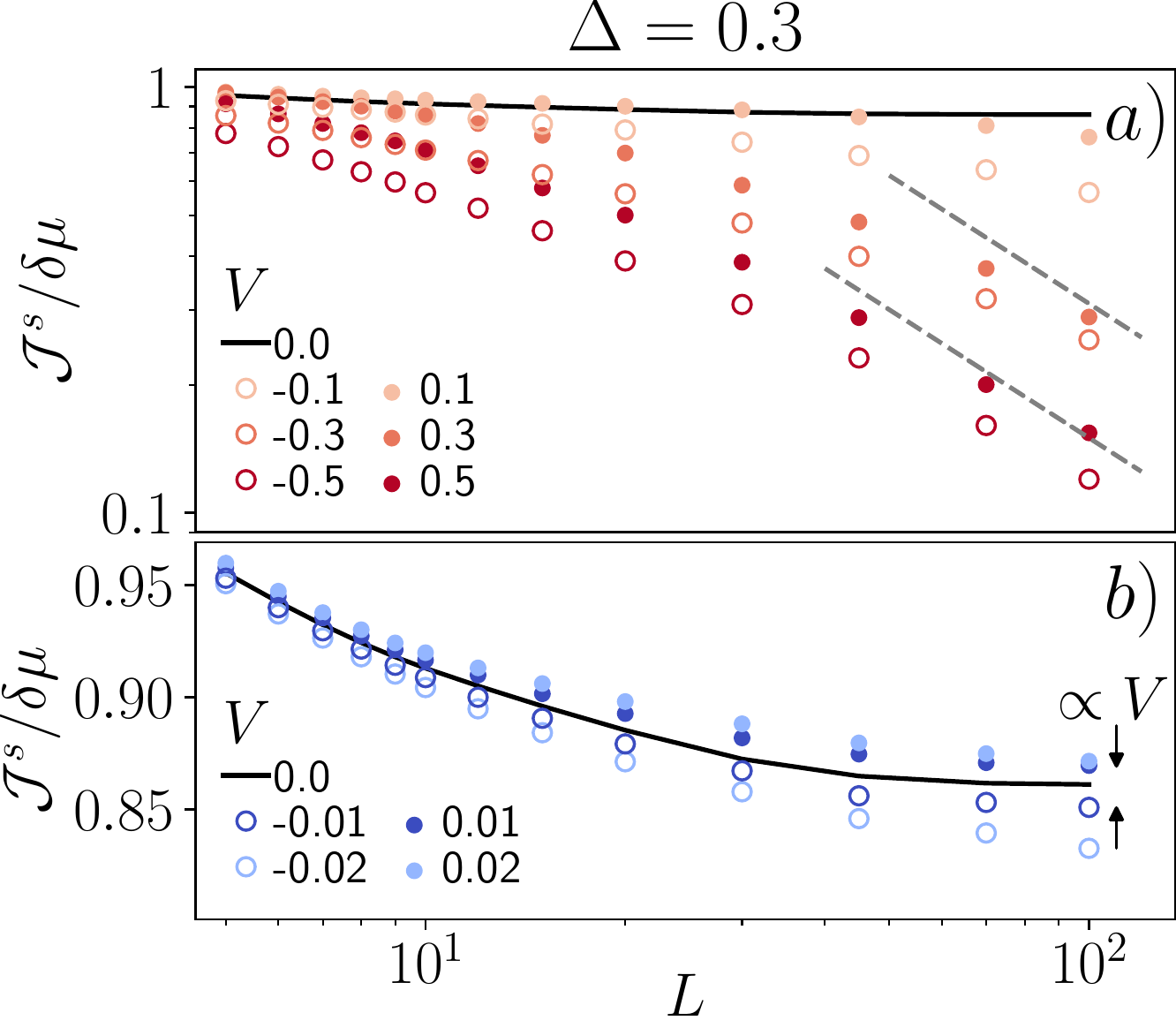} 
\par\end{centering}
\caption{\label{fig:first-3} System-size dependence of the XXZ current in the
presence of non-integrable interactions~\eqref{eq:nn} and for different IB strengths $V$. In all cases, we compare to the integrable ballistic case for $\D=0.3$ (solid black line). a)  For moderately strong IB ($|V|\geq0.1$), at short system sizes,  the stationary current is strongly sensitive to sign of the IB term $V$ before the onset of diffusion, which is signaled by the dashed-gray at larger system sizes. b) Illustration of the quasi-ballistic regime in the  $V\rightarrow0$ limit. For weak IB, the ballistic regimes appears to be just renormalized by linear (sign-dependent) corrections in $V$. }
\end{figure}

We now study the transition to the diffusive regime induced by the
IB term~\eqref{eq:nn} for $|\Delta|<1$. 
In Fig.~\ref{fig:first-3}, we present the size dependence of the spin current for $\D=0.3$ and different IB parameters. Figure~\ref{fig:first-3}a highlights the suppression of the current density for large systems and strong IB. The suppression  is compatible with a diffusive scaling, marked by the dashed grey lines. However, the observation of a clear diffusive behavior lays beyond the available system sizes. Thus we cannot conclude about the $\Delta$-dependence of the diffusion constant~\cite{Sanchez2017}.

Nevertheless, the most striking and visible effect in Fig.~\ref{fig:first-3} is not the current suppression, but rather the strong sensitivity of $\mathcal J^{s}$ to the sign of the coupling constant $V$. This dependence is visible for any system size and any IB strength and it  is absent in the non-interacting limit ($\D=0$). Two features of such phenomenon deserve particular attention:  

 {\it i)} the value of the current can even {\it increase} with respect to the integrable case after breaking integrability. This relative increase is more pronounced for small $V\ll 1$, see Fig.~\ref{fig:first-3}b, but  persists up to system sizes of the order of fifty sites for non-perturbative values of the IB strength $V\sim 0.1$, see Fig.~\ref{fig:first-3}a. This is surprising, given the expectation that IB is supposed to trigger diffusion, and thus suppress the current as function of the system size.

{\it ii)} in the limit of $V\rightarrow0$, the effects of IB appear to just renormalize the relaxation towards the ballistic regime and the saturation value of the current, see Fig.~\ref{fig:first-3}b. Breaking of integrability marks a correction to the ballistic regime, long before the scattering length $L^*$ and the related onset of diffusion. The curves for $|V|\leq 0.02$ also suggest that this effect reduces as $V\rightarrow 0$, while simultaneously persisting for larger systems. This effect has nothing to do with the boundary corrections discussed in Sec.~\ref{subsec:XX-Limit}.

This strong  sensitivity of the current to the sign of the IB term $V$ hints at the existence of linear effects in $V$ whose fate in the thermodynamic limit is intriguing. In particular, concerning the renormalization of the ballistic regime observed in Fig.~\ref{fig:first-3}b. In the next section, we argue how linear corrections control the IB crossover to the diffusive regime, giving rise to an emergent mesoscopic  ``quasi-ballistic'' regime.

\subsection{The ``quasi-ballistic'' regime}

The perturbation theory carried out in the previous sections provides some insight into the nature of the linear correction in $V$. It arises as a second-order term in $\Delta V$ when perturbing the current close to the XX limit. 
\begin{equation}\label{eq:PT-VD}
\mathcal J^s\approx\Big[1-\fl-\gl+V\Delta f_{V\Delta}(L) \Big]\delta\mu\,,
\end{equation}
Similarly to the $\D^2$ and $~V^2$ corrections, $f_{V\Delta}$ also diverges linearly with the system size $L$, see App.~\ref{sec:perturbation}. The second-order nature ($\D V$) of the linear corrections indicates that the effects discussed here only pertain the interplay between integrable and IB interactions. In the absence of nearest-neighbor interactions ($\D=0$), the effects are trivially absent, as shown in  Section~\ref{subsec:XX-Limit}.

 To understand the fate of  the linear correction in the thermodynamic limit, we  rely on a  systematic study of the finite-size scaling of the current  at finite $\D$. We numerically probe the $V\rightarrow 0$ limit by assuming a polynomial expansion of the current:
\begin{equation}\label{eq:V-only-PT}
\frac{\mathcal J^s}{\delta\mu}=1-c_{0}(\D,L)+Vc_{1}(\D,L)-V^{2}c_{2}(\D,L)+\mathcal{O}(V^{3})
\end{equation}
which extends Eq.~\eqref{eq:PT-VD} beyond the perturbative regime.

\begin{figure}
\begin{centering}
\includegraphics[width=1\columnwidth]{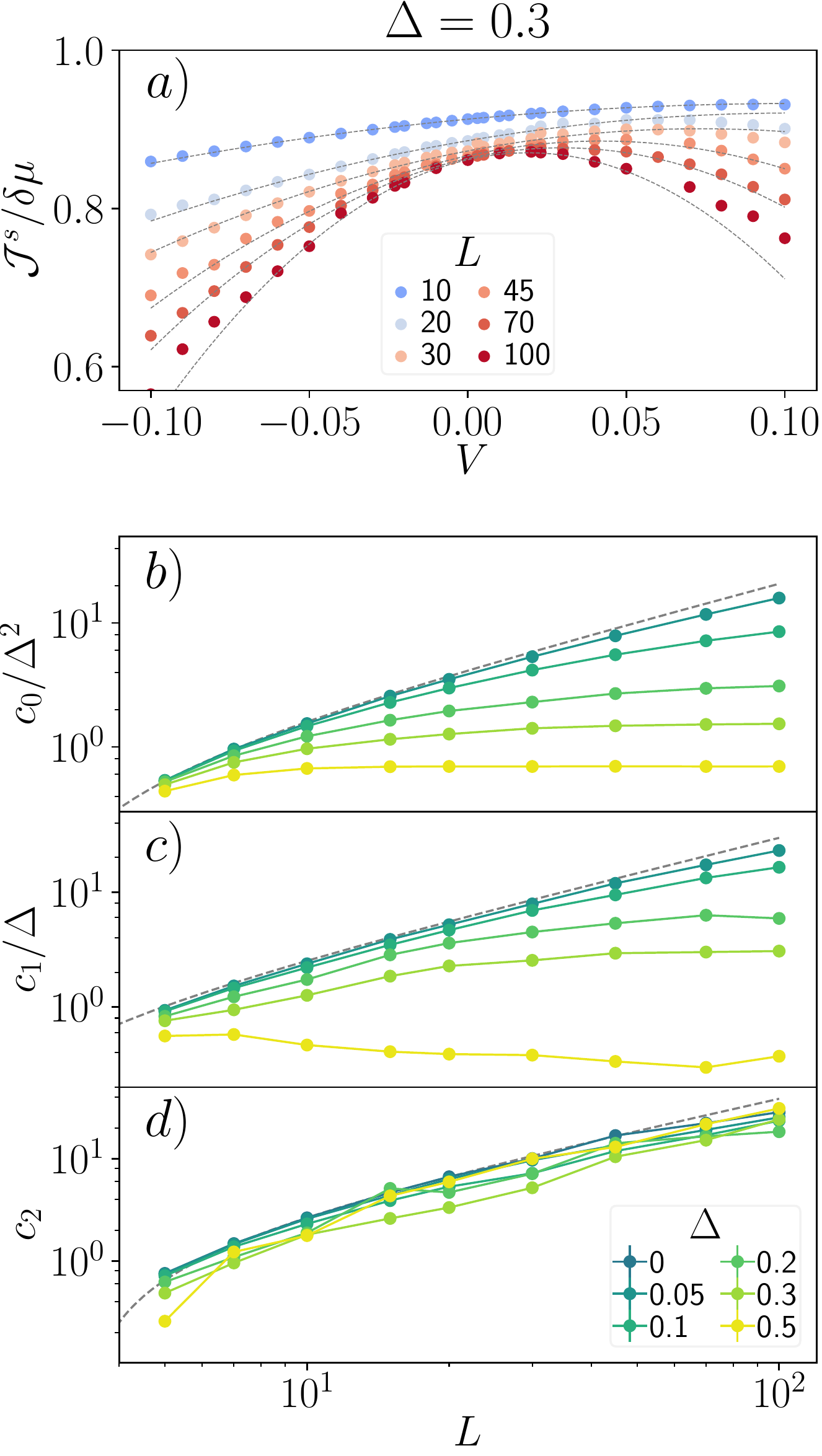} 
\par\end{centering}
\caption{a) Dependence of $\mathcal{J}^{s}$ for $\protect\D=0.3$
and different system sizes, as a function of small IB parameters. Dashed lines represent the fitting functions of Eq.~\ref{eq:V-only-PT}).
b-d) System-size dependence of the fitting parameters
for different $\D$ parameters. The dashed-gray lines correspond to the predictions from second order perturbation theory. For small $\D$, the results approach the perturbative predictions.}\label{fig:second}
\end{figure}

In Fig.~\ref{fig:second}a, we depict the dependence of the spin current $\mathcal J^{s}$ on the IB perturbation strength $V$  for finite $\D=0.3$ and increasing system sizes $L$. For $V\ll\D$, all curves can be nicely fitted with expression~\eqref{eq:V-only-PT}, with  $c_{i}$ as free parameters. The asymmetry of the parabolic dependence on $V$  is a clear indicator of the presence of linear corrections for $\D\neq0$. In Figs.~\ref{fig:second}b-d,
we show the obtained finite-size scaling of the coefficients $c_i$ for different values of $\D$. The dashed lines correspond to the analytic predictions derived with perturbation theory in Eq.~\eqref{eq:PT-VD}. They  show an excellent agreement with the numerics in the $\D\rightarrow0$ limit~\footnote{ The parameter $c_2$ deviates slightly from PT results in the $\D\rightarrow 0$. This is an artifact of the fitting procedure as argued in App.~\ref{sec:perturbation}.}. 

Figure~\ref{fig:second}c clearly shows that the coefficient $c_1$, which controls the  linear corrections in $V$, behaves analogously to $c_0$,  and thus saturates to  finite values at systems sizes of the order of $L_\D$. The finite value of $c_1$ contributes to the strong sign-dependence at moderately strong $V$ in Fig.~\ref{fig:first-3}a .

On the other hand, $c_{2}$, which controls the second-order corrections to $\mathcal J^s$ in $V$, increases linearly with $L$. Its divergence is only weakly affected by the presence of a finite $\D$. The fact that only the terms of order $\mathcal O(V^2)$ diverge suggests that
 the diffusive regime is established at the scattering length $L^*\sim1/V^{2}$ \cite{Friedman2020,Durnin2020,Moller2020}.

The different size dependence of the coefficients $c_1$ and $c_2$ corresponds to strong quantitative effects of IB on integrable systems. First of all, the non-zero linear corrections in $V$ signal that IB has prominent effects at system sizes much shorter than $L^*$. In contrast to the XX case, IB does not primarily lead to the inelastic scattering of quasi-particles and the onset of diffusion. Instead, IB leads to a transient and mesoscopic ``quasi-ballistic'' regime, in which the value of the ballistic current in the XXZ model is just renormalized by IB. Whether such corrections can be interpreted as a renormalization of the quasi-particle velocity is left for future investigation.

Additionally, such quasi-ballistic regime persists up to a novel and parametrically large length scale $L_{\D V}$,  which controls the onset of diffusion and is much shorter than $L^*$.  An estimate of $L_{\D V}$ can be obtained from Eq.~\eqref{eq:V-only-PT}. It is defined as the length scale at which  the diverging term of order $\mathcal O(V^2)$ dominates the linear correction of order $\mathcal O(V)$. For $|V|\ll\D$, we can define such length scale as 
\begin{equation}
L_{\D V}\sim\frac{ 1}{|V|c_1(\D,L\rightarrow\infty)}\ll L^*\sim\frac1{V^2}\,.
\end{equation}
This emergent length scale marks the system sizes up to which IB acts as a renormalization of the ballistic current of the integrable XXZ model with $|\D|<1$. For system sizes $L\simeq L_{\D V}$, the deviations from the ballistic regime become sizable, and the crossover to diffusion starts. Remarkably, such transient length scale does not emerge from generalized hydrodynamics
approaches~\cite{Friedman2020,Durnin2020,Moller2020}. The reason is that generalized hydrodynamics is a ``coarse-grained'' approach, which considers the limit $L\rightarrow\infty$ before $V\rightarrow0$.  Our numerical and analytical predictions rely on the opposite order of limits, which will be  relevant to study  IB  in real experiments. We expect our effects to appear on the transient time-scales controlling the quantum evolution after  quenches.

We conclude this section by stressing that the existence of such linear effects compromises the possibility to collapse the crossover to diffusion onto a unique, universal curve. 


\section{Conclusions }\label{sec:conclusions}
In this work, we studied and characterized the effects of integrability-breaking on the spin current of a boundary-driven  chain. We have first considered integrability breaking of the non-interacting XX chain. We showed that the crossover to diffusion is indeed universal and controlled by a novel scaling parameter, $V^2f_{V^2}(L)$, which we computed using perturbation theory and which accounts for boundary effects. The universal scaling found here allows to show that the ballistic-to-diffusive crossover  is controlled by the scattering length $L^*\sim V^ {-2}$, consistently with FGR. Accounting for boundary effects was important to verify the universality of such transition based  on numerical calculations. 
 
Nevertheless, the fact that deviations from ballisticity in the XX model are controlled by second order corrections in the IB strength $V$  is not trivial. In Ref.~\cite{Neuenhahn2012}, the evolution of eigenstates in the presence of IB terms was studied exactly for the same model. In that work, it was  pointed out that, for fixed system sizes $L$, perturbation theory is expected to fail for systems sizes $\tilde L\propto V^{-1/2}$. Such an estimate is readily derived by noticing that NNN interactions have typical matrix elements of order $V/L$ which couple  $\rho\propto1/L^3\ln L$ states, because of total momentum conservation. Now, it is remarkable that the length scale $\tilde L$ does not appear at all in the finite-size scaling of the current during the onset of diffusion. The physical effects of such length scales pave the way to stimulating investigations concerning other effects of integrability breaking. It is also an interesting line of investigation to extend our approach to the regimes in which $V$ is of the order of the spin exchange $J$, or larger.

We have then addressed the effects of IB in the ballistic regime of the XXZ model. Our observations are consistent with a diffusive regime in the thermodynamic limit, even though the precise determination of the diffusion constant in the presence of a finite $\D<1$ and $V\rightarrow0$ remains an interesting (and challenging) line of investigation \cite{Sanchez2017}.  Our main result, is that IB controls the ballistic-to-diffusive transition in a non-trivial way for interacting models at mesoscopic length scales. Unlike the non-interacting XX case, we showed that linear corrections in $V$ influence transport long before the onset of diffusion. This is surprising given the expectation that IB would simply suppress the current as a function of size. The fact that the opposite may happen  in the mesoscopic quasi-ballistic regime is a qualitatively new effect of interactions. As mentioned above, the physical meaning of such effect has to be clarified. 

An interesting direction would be to compare and make the connection of our findings with the time-scales describing the relaxation of non-integrable quantum systems \cite{Brenes2018,Brenes2020}. For instance, by studying the unitary evolution of a weakly polarized domain-wall state  \cite{Ljubotina2017,CastroAlvaredo2016,Bertini2016,Jesenko2011}. It would be important to understand the role of IB terms different from  Eq.~\eqref{eq:nn}, such as disorder, single impurities~ \cite{Brenes2018,Brenes2020,znidaric_weak_2020}, stochastic quantum noise~\cite{eisler_crossover_2011,bauer2017stochastic,bernard2018transport} and also dephasing~\cite{znidaric_transport_2013}.

Future research directions could address the propagation of energy and spin~\cite{MendozaArenas2019} in the presence of IB. In particular, whether the Wiedemann-Franz law~\cite{Ashcroft2003} is restored in the presence of IB terms, since it is notoriously violated in such systems at low temperatures~\cite{Kane1996,Fazio1998,Filippone2016b,Bulchandani2020}. An additional perspective is the investigation  of different integrability perturbations and their effect on quantum ladder systems attached to reservoirs~\cite{Salerno2019,Filippone2019,Greschner2019}.  


\section*{Acknowledgments}
J. S. F. and M. F. acknowledge several discussions with Dmitry A. Abanin, Vincenzo Alba, Christophe Berthod, Thierry Giamarchi, Tony Jin and, in particular, with Marko Žnidaric during the whole realization of this work and support from the FNS/SNF AmbizioneGrant PZ00P2\_174038. J.S.F acknowledges Miles Stoudenmire and Matthew Fishman for the helpful support with the ITensor library and thanks Michael Sonner and Sofia Azevedo for the careful reading of the manuscript.


\appendix


\section{Numerical details\label{sec:Numerical-details}}

Except for the large coupling limit, $V\rightarrow\inf$,
the models presented in the main text have a unique non-equilibrium steady state (NESS) in the thermodynamic
limit. This condition ensures that we can reach the NESS via a real
time-evolution $\r_{\inf}=\lim_{t\rightarrow\inf}\exp(\LL t)\rho(0)$
of any initial state $\rho(0)$. We initialize the state in the product
state $\rho(t=0)=\mathbb{I}^{\otimes L}/2^{n}$.

For small systems, $L<8$, we use exact diagonalization as baseline for other time-evolution methods. Beyond $L=8$, we
employ time-evolving block decimation (TEBD), which allows us to efficiently find the
NESS of  large spin chains, $L\lesssim100$. The algorithm was
first explored in Ref.~\cite{Prosen2009} and consists of applying
a Suzuki-Trotter decomposition of the Lindblad super-operator to the
state $\r$. In our case, we use a 4th order decomposition introduced
in Ref. \cite{Prosen2006}. At any time during the time-evolution,
the density matrix can be written in a matrix product operator form
\begin{equation}
\rho=\sum_{\{i\}} M_{1}^{i_{1}}M_{2}^{i_{2}}...M_{L}^{i_{L}} \left(\s_{1}^{i_{1}} \otimes \sigma_{2}^{i_{2}}\otimes...\otimes \s_{3}^{i_{L}} \right)
\end{equation}
where we choose the local basis to be the Pauli matrices $\s^{0,1,2}=\s^{x,y,z}$ and
$\s^{3}=\id$ and $\dim(M_{k}^{i})=\chi\times\chi$. In general, the application
of non-unitary two-site gates leads to nonphysical states as it breaks the orthogonality condition assumed in TEBD. To avoid reorthogonalizing the MPO at every time-step,
we apply the gates sequentially instead~\cite{Daley2005}. We simulate
the next-to-nearest interaction using the swap-gate technique.

In the presence of interactions, the necessary bond dimension $\chi$
to simulate the NESS is expected to grow with the system size. We consider
that the time evolved state $\r(t)$ correctly approximates the NESS
if it satisfies three criteria: the current is homogeneous across
the chain, the average current does not evolve in time and the current
converges as the bond-dimension increases.  Next, we present the algorithm used in this paper. The quantity  $\bar{\mathcal J}$
represents the spatial average of the spin current
\begin{enumerate}
\item Initialize with the product state $\rho(t=0)=\mathbb{I}^{\otimes L}/2^{n}$
($\chi_{0}=1$)
\item Increase the bond-dimension by $\chi_{i}=\delta\chi+\chi_{i-1}$.
\item Time-evolve the state until the current has saturated in time.
\begin{enumerate}
\item Compute the time variance in the last $T=10,30$ time units (of hopping)
$\s_{T}^{2}=\sum_{i=1}^{T}(\bar{\mathcal J}(t-i)-\mu)^{2}/T$
\item Repeat step 3 until $\left|\s_{30}^{2}-\s_{10}^{2}\right|/\s_{30}^{2}<1\%$
\end{enumerate}
\item Check convergence
\begin{enumerate}
\item Compute the spacial variance $\s^{2}(\mathcal J)=\sum_{i=2}^{L-1}(\mathcal{J}_{i}-\bar{\mathcal J})^{2}/(L-2)$
\item Compute the change with the bond dimension $\e_{\x}=\bar{\mathcal J}(\chi_{i})-\bar{\mathcal J}(\chi_{i-1})$
\item Repeat steps 2,3 and 4 until $\left|\s(\mathcal J)/\bar{\mathcal J}\right|<1\%$ and
$\left|\e_{\chi}/\bar{\mathcal J}\right|<0.5\%$
\end{enumerate}
\item Compute the final current and associated error $\e_{J}=\max(\s(\mathcal J),\e_{\chi})$.
\end{enumerate}
In most situations, we require a must stricter bond on the homogeneity
condition, often requiring $\left|\s(\mathcal J)/\bar{\mathcal J}\right|<0.1\%$. 
The time step of the Trotter decomposition is variable along the algorithm. For small bond dimensions, we use a large time step, $dt=0.2$, to quickly advance the simulation and reduce it when closer to convergence, up to $dt=0.05$.
Due to the  convergence criterion employed, simulations can take weeks to converge or reach inaccessible bond dimensions. For this reason, if the criterion are not satisfied for $\chi\leq160$, we consider that the system has not converged and do not show it.

The algorithm was implemented using the open-source ITensor library \cite{itensor}.


\section{Diagonalization of the XX chain\label{app:3quant} \label{subsec:appedix-eigendecomp}}

In this section, we provide a summary on how to diagonalize the non-interacting XX limit, $V=\D=0$. We follow the protocol of
Ref.~\cite{Guo2017} which reduces the diagonalization problem to
finding the eigenbasis of a $2L\times2L$ matrix. It is useful to
work in the fermionic representation via the Jordan-Wigner transformation

\begin{equation}
\begin{split}
\s_{j}^{+} & =e^{-i\pi\sum^{j-1}n_{k}}a_{j}^{\dag}\\
\s_{j}^{-} & =e^{i\pi\sum^{j-1}n_{k}}a_{j}\\
\sigma_{j}^{z} & =2a_{j}^{\dag}a_{j}-1
\end{split}
\end{equation}

In the fermionic representation, the Hamiltonian becomes
\begin{equation}
H_{\rm XX}=\sum_{i,j=1}^{L}h_{ij}a_{i}^{\dag}a_{j}
\end{equation}
with $h_{i,j}=2J\delta_{|i-j|,1}$. Since the $\{a_{i},a_{i}^{\dag}\}$
operators act left and right of the density matrix, it is useful to
work in the Liouville space of super-operators. In the super-operator formalism,
density matrices are mapped onto vectors in a vector space of dimensions
$\mathbb{C}^{4^{L}}\times\mathbb{C}^{4^{L}}$ according to the mapping
$\kket{M\r N}=M\otimes N^{T}\kket{\rho}$, where $\kket{\rho}$ is
the row-vectorized form of the matrix $\rho$. We can now define a
new set of $2L$ super-operators $\mathcal{B}=\left\{ \hat{b}_{i},\hat{b}_{i}^{\dag},\hat{b}_{L+i},\hat{b}_{L+i}^{\dag}\right\} _{i=1}^{L}$
which act on $\kket{\rho}$ according to:
\begin{equation}
\begin{split}
\hat{b}_{i}\kket{\r} & =\kket{a_{i}\rho}\\
\hat{b}_{i}^{\dag}\kket{\r} & =\kket{a_{i}^{\dag}\rho}\\
\hat{b}_{L+i}\kket{\r} & =\kket{\mathcal{\hat{P}}(\rho a_{i}^{\dag})}\\
\hat{b}_{L+i}^{\dag}\kket{\r} & =\kket{\mathcal{\hat{P}}(\rho)a_{i}}
\end{split}
\end{equation}
where $\mathcal{\hat{P}}=e^{i\pi\sum n_{i}\otimes\id+\id\otimes n_{i}^{T}}$
is a super-operator string which imposes the necessary anti-commutations
relations $\{\hat{b}_{i}^{\dag},\hat{b}_{j}\}=\delta_{ij}$. In practice, the
$\mathcal{B}$ basis acts as a complete set of creation and destruction
operators in the occupation number basis of a lattice of size $2L$.
Physically, $\mathcal{\hat{P}}$ is a parity operator with eigenvalues $\pm1$
and counts the number of excitations in the new fermionic system with
$2L$ states. For reasons clear bellow, we will only be interested
in $\mathcal{\hat{P}}=1$. In the new $\mathcal{B}$ basis, the Lindblad super-operator reads:
\begin{equation}
\begin{split}
\LL_{\rm XX} & =-i\sum_{i,j=1}^{L}\left(h_{i,j}\hat{b}_{i}^{\dag}\hat{b}_{j}-h_{ji}\hat{b}_{L+i}^{\dag}\hat{b}_{L+j}\right)\\
 & +\sum_{i=1,L}\G(1+\mu_{i})\left(2\hat{b}_{i}^{\dag}\hat{b}_{L+i}^{\dag}\mathcal{\hat{P}}-\hat{b}_{L+i}\hat{b}_{L+i}^{\dagger}-\hat{b}_{i}\hat{b}_{i}^{\dagger}\right) \\
& +\sum_{i=1,L}\G(1-\mu_{i})\left(-2\hat{b}_{i}\hat{b}_{L+i}\mathcal{\hat{P}}-\hat{b}_{L+i}^{\dag}\hat{b}_{L+i}-\hat{b}_{i}^{\dag}\hat{b}_{i}\right)
\end{split}
\end{equation}

Similarly to the diagonalization procedure of quadratic Hamiltonians, we are interested
in finding a basis of $2L$ creation and annihilation super-operators $\mathcal{C}=\left\{ \c'_{i},\c_{i}\right\} _{i=1}^{2L}$
that diagonalizes the unperturbed problem, $\hat{\LL}_{\text{XX}}=\sum_{i=1}^{2L}\a_{i}\c'_{i}\c_{i}$. If such basis exists, the eigenstates of $\hat{\LL}_{\text{XX}}$
can be constructed from excitations on the vacuum state of the $c$'s
operators, $\kket{\rho_{\mu}}=\sum_{\left\{ \mu_{i}\right\} }\c'_{1}{}^{\mu_{1}}...\c'_{2L}{}^{\mu_{2L}}\kket 0$ and $\l_{\mu}=\sum_{i}^{L}(\mu_{i}\a_{i}+\mu_{L+i}\a_{i}^{*})$. Trivially, the NESS is the vacuum state of the $\mathcal{C}$ basis.

Due to particle hole symmetry, the values of $\a_i$ must come in conjugate
pairs $\{\a,\a^{*}\}$ with $\Re(\a)\leq0$. We fix $\a_{i}^{*}=\a_{L+i}$
in our notation. In general, the Lindblad super-operator is not hermitian
and neither are the $\c$'s super-operators, however they still respect
the fermionic anti-commutation relations $\{\c_{i},\c'_{j}\}=\delta_{i,j}$
and $\{\c_{i},\c{}_{j}\}=\{\c'_{i},\c'{}_{j}\}=0$.
The $\c,\c'$ operators represent a linear super position of particle
and hole excitations acting both left and right of the density matrix
and should be understood as the ``normal modes'' of the open system.
The exact mapping between $\{\c_{i},\c_{i}'\}$ and $\{\hat{b}_{i},\hat{b}_{i}^{\dag}\}$
operators can be found in Ref.\cite{Guo2017} and shown here for completeness

\begin{equation}
\left[\begin{array}{c}
\hat{b}_{1\rightarrow L}\\
\hat{b}_{L+1\rightarrow2L}^{\dagger}\\
\hat{b}_{1\rightarrow L}^{\dagger}\\
\hat{b}_{L+1\rightarrow2L}
\end{array}\right]=\left[\begin{array}{cc}
W & 0\\
0 & -Y_{L}W^{*}Y_{L}
\end{array}\right]\left[\begin{array}{c}
\c_{1\rightarrow L}\\
\c'_{L+1\rightarrow2L}\\
\c'_{1\rightarrow L}\\
\c_{L+1\rightarrow2L}
\end{array}\right]\label{eq:B2C-1}
\end{equation}
where $Y_{L}=-i\left[\begin{array}{cc}
0 & \id_{L}\\
-\id_{L} & 0
\end{array}\right]$ and the columns of $W$ are the right eigenvectors of a matrix
$M$.
In our work, $M$ acquires a simple form 
\begin{equation}
M=\frac{1}{2}\left[\begin{array}{cc}
-ih+\L^{+}-\L^{-} & 2\L^{+}\\
2\L^{-} & -ih-\L^{+}+\L^{-}
\end{array}\right]
\end{equation}
 with diagonal matrices $\L_{i}^{+}=(\d_{i,1}+\d_{i,L})\G(1+\mu_{i})$
and $\L_{i}^{-}=(\d_{i,1}+\d_{i,L})\G(1-\mu_{i})$.

\begin{figure}[t]
\begin{centering}
\includegraphics[width=1\columnwidth]{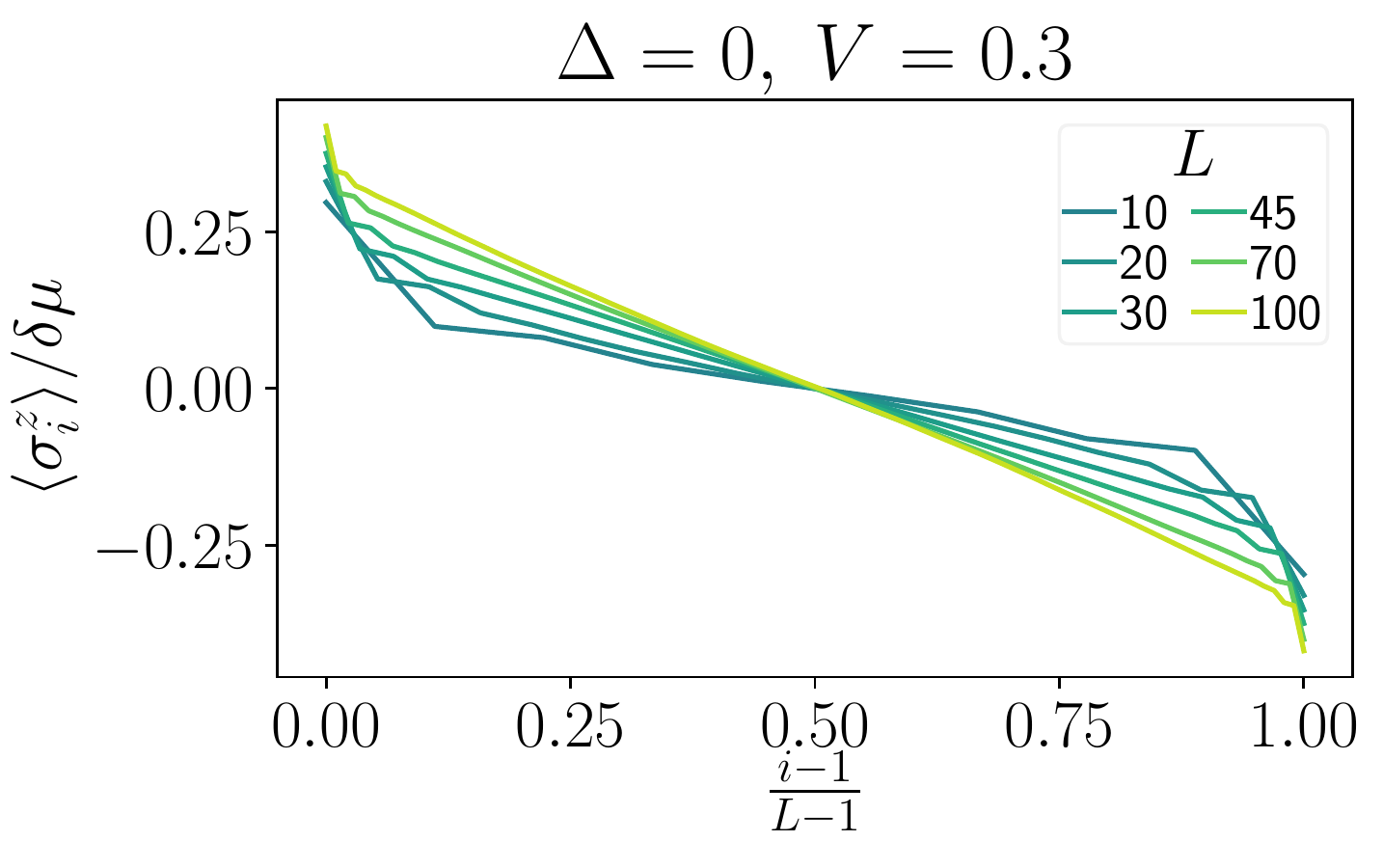}
\includegraphics[width=1\columnwidth]{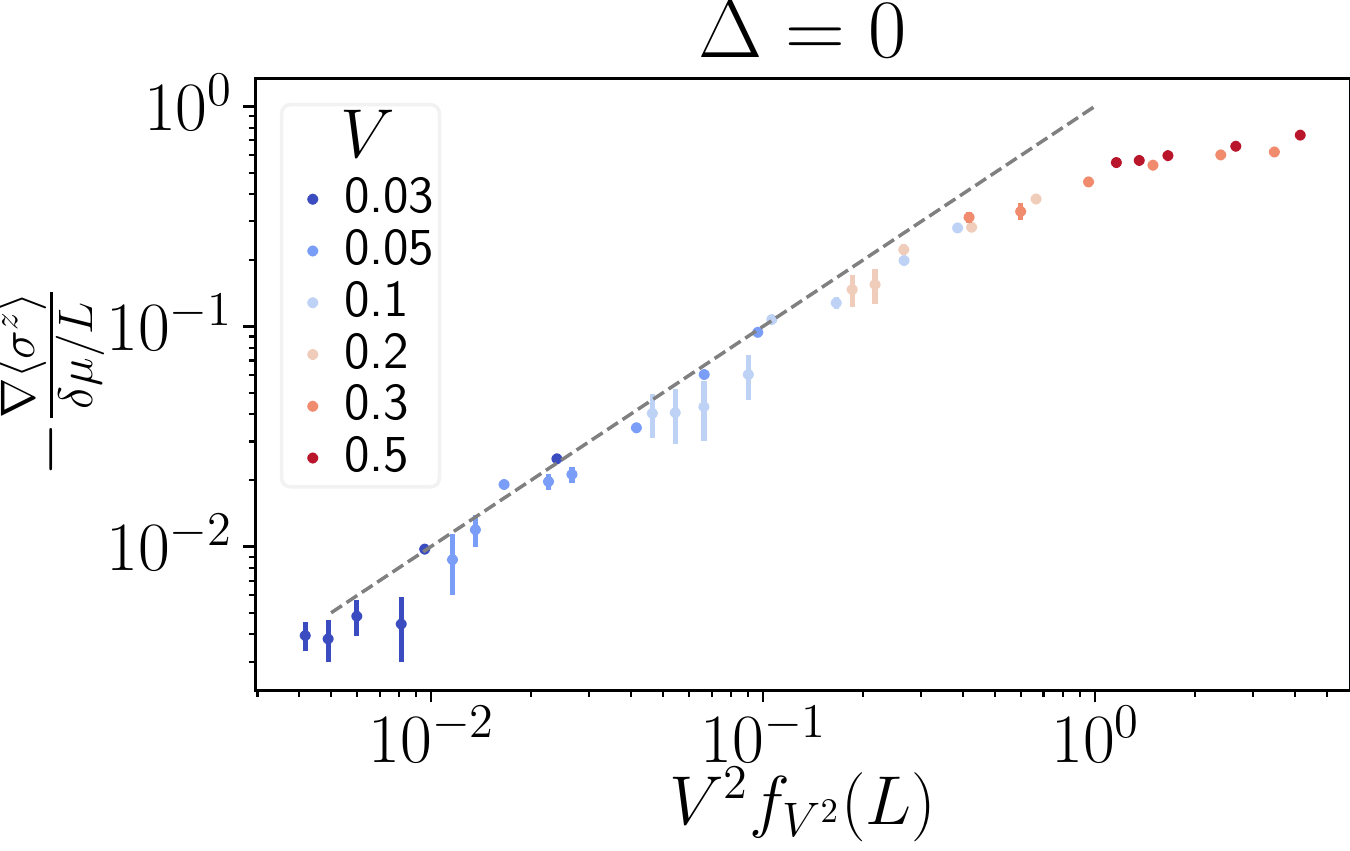}
\par\end{centering}
\caption{\label{fig:mag} Top) Magnetization profile close to the diffusive regime for different system sizes. Bottom) Rescaled slope of the magnetization for different IB parameters as a function of $\fl$. System sizes range from $L=15$ to $L=100$. Analytic predictions close to the ballistic regime are depicted as dashed lines. In the diffusive regime ($V^2f_{V^2}(L)\approx1$), the magnetization slope approaches $\delta\mu/L$ as expected.}
\end{figure}

To our knowledge, there is no analytical solution for $W$ as a function of $L$ and so we resort to exact diagonalization. Once the mapping of Eq. \eqref{eq:B2C-1} is found, we can express any super-operator in the $\mathcal{C}$ basis.


\section{Universal Scaling}\label{app:scaling}

In this Appendix, we provide further details on the universality of the scaling discussed in Sec. \ref{subsec:XX-Limit}.

It follows from Fick's law that, when imposing a fixed bias, the magnetization profile interpolates linearly between the borders. However, this is only true in the thermodynamic limit, and finite systems present small deviations up to four sites into the chain's bulk. In Fig. \ref{fig:mag}-top, we depict the magnetization profile of the XX model close to the diffusive regime for different system sizes.  The effects of the border are visible up to very large  systems, $L=100$.

For consistency, we verify that the magnetization's gradient converges to $\nabla \av{\sigma^z}=-\delta \mu/L$ in the diffusive regime, $\fl \gg 1$, see Fig. \ref{fig:mag}-bottom. There, we depict the rescaled gradient of $\sigma^z$ obtained by a linear fit of the magnetization close to the middle of the chain. We find an overall scaling with $\fl$ but, in contrast to Fig. \ref{fig:first}, the  finite size effects in the magnetization profile lead to non-negligible deviations. Close to the ballistic regime, we find a moderate agreement with $\nabla \av{\sigma^z}=-\frac{\delta \mu}{L} \fl$, depicted as dashed gray line.


\begin{figure}
\begin{centering}
\includegraphics[width=.9\columnwidth]{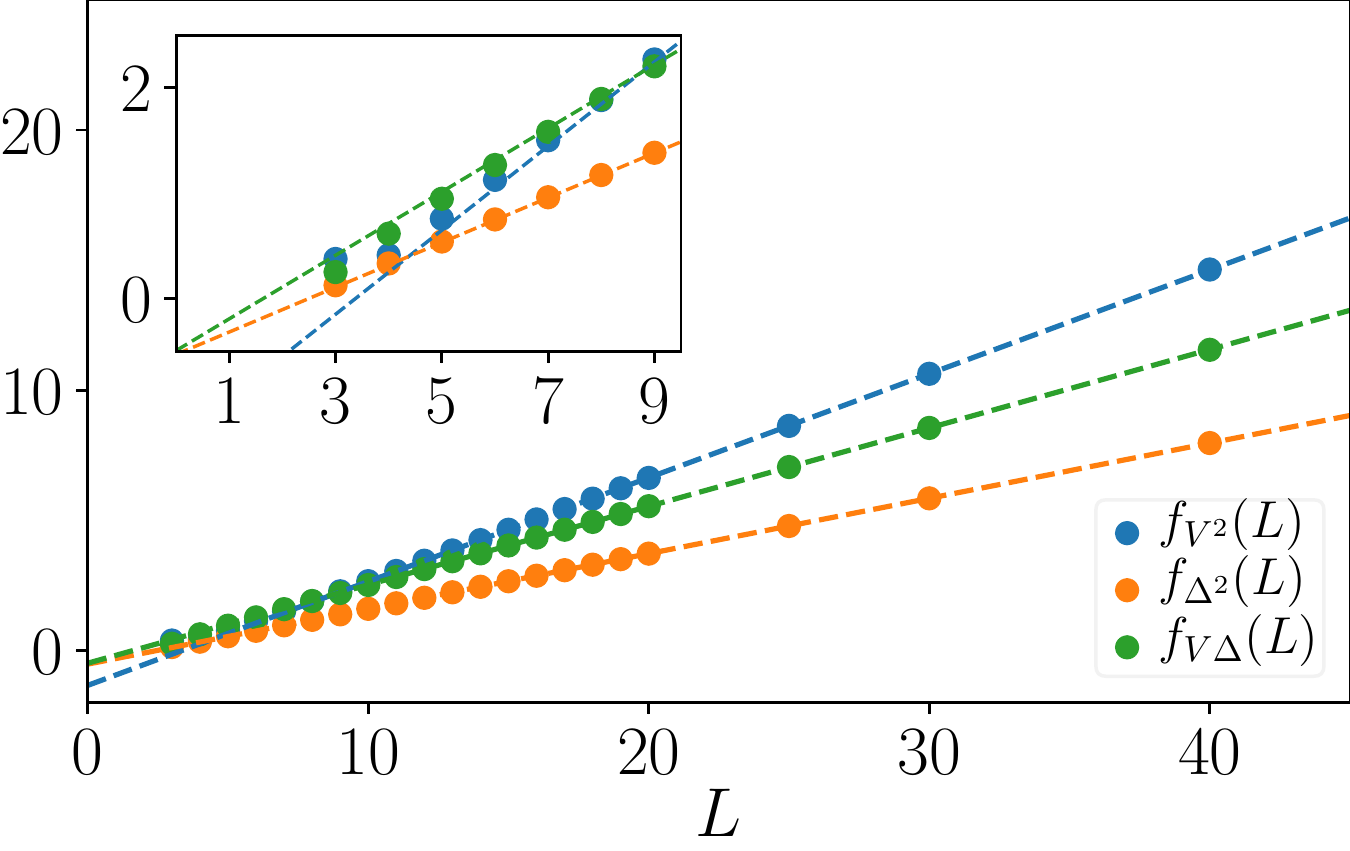}
\par\end{centering}
\caption{\label{fig:fi} System size dependence of the functions $f_i$ in Eq. \eqref{eq:PT}. Dashed lines depict the linear fitting performed beyond $L=6$.
Inset: highlight for very short systems, in which the deviations from perfect linear scaling can be appreciated.}
\end{figure}

\section{Perturbation Theory}\label{sec:perturbation}

As mentioned in the main text, perturbation theory (PT) provides a benchmark and helpful insights on the numerical data in the limit of small interactions. In this section, we provide further details on the method.

The object of interest is the NESS of the system. It corresponds to the unique (in our case) zero eigenvalue of the non-unitary master equation~\eqref{eq:master-eq}. This equation can be written in terms of the Liouvillian super-operator $d_t\rho=\LL (\rho)$. Super-operators are denoted by a hat.

The first step in PT is to find the eigendecomposition of the unperturbed problem, i.e. the super-operator of the non-interacting boundary-driven XX model, $\LL_{\rm XX}$. As a direct consequence of the non-unitarity of general Lindblad evolutions, the Lindblad super-operator is described by a non-hermitian matrix and thus has different
left and right eigenvectors, $\tilde{\r}_{\mu}$ and $\r_{\mu}$ respectively. They respect
the normalization condition $\Tr (\tilde{\r}_{\mu} \r_{\nu})=\delta_{\mu\nu}$  and share the same eigenvalue $\l_{\mu}$, whose real part corresponds to the physical relaxation rate of $\r_{\mu}$.

The eigenstates of $\LL_{\rm XX}$  serve as the basis to perturbatively construct the eigenstates of the full problem. 
Since $\LL_{\rm XX}$ is a quadratic super-operator, it is useful to rely on the third-quantization formalism~\cite{Prosen2008,Guo2017} to find its eigendecomposition. In Sec.~\ref{subsec:appedix-eigendecomp}, we construct the $4^L$ eigenstates $\r_{\mu}$
by consecutively acting with annihilation(creation) operators, $\hat{c}^{(')}$ on a vacuum state of $2L$ particles, $\rho_0$. This approach allows to diagonalize the Lindblad super-operator, which can be written as 
\begin{equation}
\begin{split}
\LL_{\text{XX}}(\circ)&=\sum_i^{2L}\alpha_i\hat{c}'_i\hat{c}_i(\circ)\\
&=\sum_{\mu}^{4^{L}}\l_{\mu} \r_{\mu}\Tr(\tilde{\r}_{\mu}  \circ)\,,
\end{split}
\end{equation}
where $\rho_{\mu}=\sum_{\left\{ \mu_{i}\right\} }\hat{c}'_{1}{}^{\mu_{1}}...\c'_{2L}{}^{\mu_{2L}}(\r_0)$ and $\l_{\mu}=\sum_{i}^{2L}\mu_{i}\a_{i}$.
All the models discussed here have a unique NESS that satisfies $\l_{0}=0$ and $\tilde{\r}_{0}=\id$. The NESS of the XX model carries a finite current proportional to the bias, $\Tr[\hat{\mathcal J^s}\rho_0]=\d \mu$. $\hat{\mathcal J^s}$ is the spin current density super-operator.

\begin{table}[b]
\begin{centering}
\begin{tabular}{|c|c|c|}
\hline 
 & $a$ & \textbf{$b$}\tabularnewline
\hline 
\hline 
$f_{V^{2}}$ &  0.3992& -1.348\tabularnewline
\hline 
$f_{\D^{2}}$ & 0.2124 & -0.5307\tabularnewline
\hline 
$f_{V\D}$ & 0.3015 & -0.4946\tabularnewline
\hline 
\end{tabular}
\par\end{centering}
\caption{Fitting parameters of the second order corrections to the current. }\label{tab:fs}
\end{table}

\begin{figure*}[th!!]
\begin{centering}
\includegraphics[width=1.9\columnwidth]{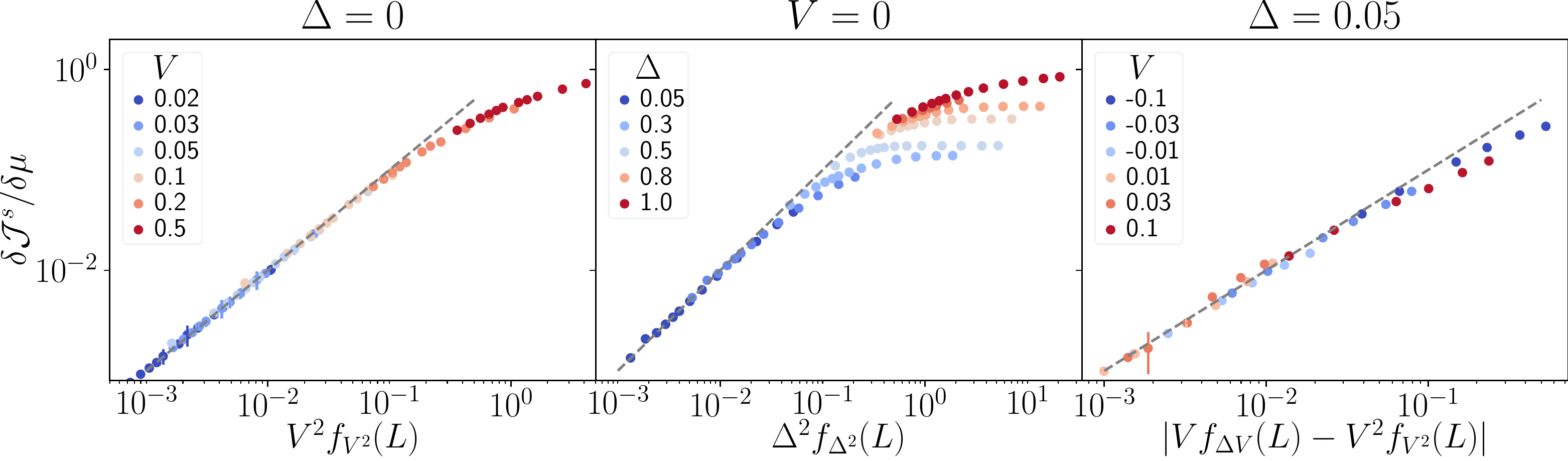}
\par\end{centering}
\caption{\label{fig:corr} Deviations from the ballistic current for small
(next-to)-nearest neighbor interactions. System sizes range from $L=5$
to $L=100$. The numerical data closely follows the analytic predictions of~\eqref{eq:PT} depicted as dashed lines.}
\end{figure*}

In the second step of PT, we  look for a perturbative solution to the NESS of Eq.~\eqref{eq:master-eq}, in the form $\rho_{ss}=\sum_{m,n=0}^{\infty}V^{m}\D^{n}\r^{(m,n)}$, where $\rho^{(0,0)}=\rho_{0}$ is the NESS of the XX model. 
Assuming orthonormality of left and right
eigenvectors, the expansion terms can be computed order by order~\cite{Li2014}:
\begin{align}
\begin{split} \label{order-exp}
V^{m}\D^{n}\rho_{m,n}=i\LL_{\text{\rm XX}}^{+}\Big(&\big[H_{\text{NNN}},\r^{(m-1,n)}\big] \\
&+\big[H^{J=0}_{\text{XXZ}},\r^{(m,n-1)}\big]\Big)\,,
\end{split}
\end{align}
where we introduced  the Moore-Penrose pseudoinverse of the super-operator $\LL_{\rm XX }(\circ)$, $\LL_{\text{\rm XX}}^{+}(\circ)=\sum_{\mu>0}\l_{\mu}^{-1}\r_{\mu}\Tr(\tilde{\r}_{\mu}\circ)$. The above perturbation scheme
ensures that at any truncation order, the density matrix remains Hermitian,
positive-semidefinite and with trace equal to one~\footnote{This is not true in general and comes from the fact that we perturb by adding a Hamiltonian term}. Equation~\eqref{order-exp} leads to the same results of the variational  approach in Ref.~\cite{Znidaric2011}.

We analytically compute corrections to the bulk's spin current up
to second order in perturbation theory. All the results are valid only in the linear response regime, $|\d \mu|\ll 1$ and discard higher order corrections $\mathcal{O}(\mu^2)$.
Assuming $\gamma=J=1$ in Eq.~\eqref{eq:master-eq}, we obtain
\begin{equation}\label{eq:PT}
\begin{split}
\mathcal J^s&=\sum_{m,n=0}^{\infty}V^{m}\D^{n}\Tr \left(\hat{\mathcal{J}_s}\r^{(m,n)} \right) \\
&\approx\Big[1-V^{2}f_{V^{2}}(L)-\D^{2}f_{\D^{2}}(L)+V\D f_{V\D}(L)\Big]\d\mu
\end{split}\,.
\end{equation}
The system size dependence of the functions $f_{V^2,\D^2,V\D}$ is shown in Fig.~\ref{fig:fi}. Beyond $L\sim5$, the scaling for all $f_{i}$ is linear in $L$ and the fitting functions  $f_{i}=a_{i}L+b_{i}$ are depicted in corresponding dashed lines and reported in Table~\ref{tab:fs}.

Notice the linearity in $V$ of the third term
in Eq.~\eqref{eq:PT}, which is responsible for the current enhancement.
It is clear from Eq.~\eqref{eq:PT} that the large $L$ limit and
the small interactions limit do not commute. For instance, both the
integral and non-integrable corrections to the XX model lead to
divergent contributions which do not capture the enforcement of ballistic
or diffusive behavior at large system sizes. 

We illustrate now the agreement with  PT and our tDMRG simulations. In the main text, we compared the PT results against a polynomial fit of the current, see Fig. \ref{fig:second}. We argued that Eq.~\eqref{eq:PT} correctly predicted the current in the limit of $\D\rightarrow0$ but some small deviations were observed in the order $\mathcal O(V^2)$ term. In Fig.~\ref{fig:corr}, we present a complementary analysis of the data which does not rely on fitting polynomials. Fig. \ref{fig:corr} depicts the correction to the current, $\delta \mathcal{J}^s$, upon turning on interactions, respectively $V$, $\Delta$ and $V$, for left, center and right plots. The $x$-axis is rescaled according to \eqref{eq:PT} and dashed gray lines depict the perturbation theory predictions. 

We can observe that, for small interactions ($\Delta$ and $V$),
the current is indeed well described by Eq.~\eqref{eq:PT}. As noted in the main text, the presence of
a single next-to-nearest neighbor interaction is characterized by
a strong scaling of the current with the variable $\fl$, Fig.~\ref{fig:corr}-left.
This is
qualitatively different from the nearest neighbor interactions where
the current converges to a value independent of $L$ and a scaling
with $\gl$ is never possible, see Fig.~\ref{fig:corr}-center. Nevertheless, we can observe that for small $\D\leq0.3$ an approximate scaling with $\gl$ might be possible. In that situation, the current would saturate after a length of order $L_{\D}\sim1/\D^{2}$. For stronger interactions, $L_{\D}$  appears to diverge close to $\D=1$ but for $0.8>\D>0.3$ the current still saturates before $\gl\lesssim10$.
Fig. \ref{fig:corr}-right shows that, for small $\D$, perturbation
theory becomes exact and that the derivations seen in Fig.  \ref{fig:second}
are indeed an artifact of the fitting.

\bibliographystyle{apsrev4-1}
\bibliography{main}

\end{document}